\PassOptionsToPackage{unicode}{hyperref}
\PassOptionsToPackage{hyphens}{url}
\PassOptionsToPackage{dvipsnames,svgnames,x11names}{xcolor}
\documentclass[
]{article}
\usepackage{amsmath,amssymb}
\usepackage{lmodern}
\usepackage{iftex}
\ifPDFTeX
  \usepackage[T1]{fontenc}
  \usepackage[utf8]{inputenc}
  \usepackage{textcomp} % provide euro and other symbols
\else % if luatex or xetex
  \usepackage{unicode-math}
  \defaultfontfeatures{Scale=MatchLowercase}
  \defaultfontfeatures[\rmfamily]{Ligatures=TeX,Scale=1}
\fi
\IfFileExists{upquote.sty}{\usepackage{upquote}}{}
\IfFileExists{microtype.sty}{% use microtype if available
  \usepackage[]{microtype}
  \UseMicrotypeSet[protrusion]{basicmath} % disable protrusion for tt fonts
}{}
\makeatletter
\@ifundefined{KOMAClassName}{% if non-KOMA class
  \IfFileExists{parskip.sty}{%
    \usepackage{parskip}
  }{% else
    \setlength{\parindent}{0pt}
    \setlength{\parskip}{6pt plus 2pt minus 1pt}}
}{% if KOMA class
  \KOMAoptions{parskip=half}}
\makeatother
\usepackage{xcolor}
\usepackage{color}
\usepackage{fancyvrb}

\DefineVerbatimEnvironment{Highlighting}{Verbatim}{commandchars=\\\{\}}
\newenvironment{Shaded}{}{}

\newcommand{\BuiltInTok}[1]{\textcolor[rgb]{0.00,0.50,0.00}{#1}}

\newcommand{\CommentTok}[1]{\textcolor[rgb]{0.38,0.63,0.69}{\textit{#1}}}

\newcommand{\ControlFlowTok}[1]{\textcolor[rgb]{0.00,0.44,0.13}{\textbf{#1}}}

\newcommand{\DecValTok}[1]{\textcolor[rgb]{0.25,0.63,0.44}{#1}}

\newcommand{\FloatTok}[1]{\textcolor[rgb]{0.25,0.63,0.44}{#1}}

\newcommand{\ImportTok}[1]{\textcolor[rgb]{0.00,0.50,0.00}{\textbf{#1}}}

\newcommand{\KeywordTok}[1]{\textcolor[rgb]{0.00,0.44,0.13}{\textbf{#1}}}
\newcommand{\NormalTok}[1]{#1}
\newcommand{\OperatorTok}[1]{\textcolor[rgb]{0.40,0.40,0.40}{#1}}

\newcommand{\SpecialCharTok}[1]{\textcolor[rgb]{0.25,0.44,0.63}{#1}}

\newcommand{\StringTok}[1]{\textcolor[rgb]{0.25,0.44,0.63}{#1}}

\providecommand{\tightlist}{%
  \setlength{\itemsep}{0pt}\setlength{\parskip}{0pt}}
\NewDocumentCommand\citeproctext{}{}
\NewDocumentCommand\citeproc{mm}{%
  \begingroup\def\citeproctext{#2}\cite{#1}\endgroup}
\makeatletter
 \let\@cite@ofmt\@firstofone
 \def\@biblabel#1{}
 \def\@cite#1#2{{#1\if@tempswa , #2\fi}}
\makeatother
\newlength{\cslhangindent}
\newlength{\csllabelwidth}
\newenvironment{CSLReferences}[2] % #1 hanging-indent, #2 entry-spacing
 {\begin{list}{}{%
  \setlength{\itemindent}{0pt}
  \setlength{\leftmargin}{0pt}
  \setlength{\parsep}{0pt}
  \ifodd #1
   \setlength{\leftmargin}{\cslhangindent}
   \setlength{\itemindent}{-1\cslhangindent}
  \fi
  \setlength{\itemsep}{#2\baselineskip}}}
 {\end{list}}
\usepackage{calc}

\ifLuaTeX
\usepackage[bidi=basic]{babel}
\else
\usepackage[bidi=default]{babel}
\fi
\babelprovide[main,import]{american}
\def\languageshorthands#1{}
\ifLuaTeX
  \usepackage{selnolig}  % disable illegal ligatures
\fi
\IfFileExists{bookmark.sty}{\usepackage{bookmark}}{\usepackage{hyperref}}
\IfFileExists{xurl.sty}{\usepackage{xurl}}{} % add URL line breaks if available
\hypersetup{
  pdftitle={ESPResSo++: A Fast and Extensible Molecular Simulation Package
for Coarse-Grained Models},
  pdfauthor={Zhen-Hao Xu, James Vance, Nikita Tretyakov, Sebastian Eibl,
Pavel Kus, Jakub Krajniak, Tristan Bereau, Horacio V. Guzman, Bin Song,
Markus Rampp, Torsten Stuehn, Christoph Junghans},
  pdflang={en-US},
  colorlinks=true,
  linkcolor={Maroon},
  filecolor={Maroon},
  citecolor={Blue},
  urlcolor={Blue},
  pdfcreator={LaTeX via pandoc}}

\title{ESPResSo++: A Fast and Extensible Molecular Simulation
Package for Coarse-Grained Models}

\definecolor{c53baa1}{RGB}{83,186,161}
\definecolor{c202826}{RGB}{32,40,38}

\usepackage[affil-it]{authblk}
\usepackage{orcidlink}
\author[4%
  ]{Zhen-Hao Xu%
    }
\author[4%
  ]{James Vance%
    \,\orcidlink{0000-0001-7112-0382}\,%
    }
\author[4%
  ]{Nikita Tretyakov%
    }
\author[2%
  ]{Sebastian Eibl%
    \,\orcidlink{0000-0002-1069-2720}\,%
    }
\author[2%
  ]{Pavel Kus%
    }
\author[6%
  ]{Jakub Krajniak%
    \,\orcidlink{0000-0001-9372-6975}\,%
    }
\author[5%
  ]{Tristan Bereau%
    \,\orcidlink{0000-0001-9945-1271}\,%
    }
\author[7%
  ]{Horacio V. Guzman%
    \,\orcidlink{0000-0003-2564-3005}\,%
    }
\author[1%
  ]{Bin Song%
    \,\orcidlink{0000-0003-4229-9242}\,%
    }
\author[2%
  ]{Markus Rampp%
    \,\orcidlink{0000-0001-8177-8698}\,%
    }
\author[1%
  ]{Torsten Stuehn%
    \,\orcidlink{0009-0006-2144-2002}\,%
    }
\author[3%
  ]{Christoph Junghans%
    \,\orcidlink{0000-0003-0925-1458}\,%
    }

\affil[1]{Max Planck Institute for Polymer Research, Mainz, Germany%
  }
\affil[2]{Max Planck Computing and Data Facility, Garching, Germany%
  }
\affil[3]{Los Alamos National Laboratory, Los Alamos, USA%
  }
\affil[4]{Johannes Gutenberg University of Mainz, Mainz, Germany%
  }
\affil[5]{Heidelberg University, Heidelberg, Germany%
  }
\affil[6]{Independent researcher, Poznań, Poland%
  }
\affil[7]{Institut de Ciència de Materials, Barcelona, Spain%
  }
\date{2 October 2025}

\begin{document}
\maketitle

\section{Summary}\label{summary}

\textbf{ESPResSo\texttt{++}} is an open-source software package for
molecular dynamics (MD) simulations with a particular emphasis on
coarse-grained (CG) models of soft matter systems
(\citeproc{ref-Praprotnik2008}{Praprotnik et al., 2008}). Written in
C\texttt{++} with a flexible Python interface, it is designed for
high-performance computing (HPC) environments and supports massively
parallel simulations through MPI. The package enables simulations of
polymers, membranes, colloids and complex fluids with a wide range of
interaction models and advanced algorithms.

ESPResSo\texttt{++} builds upon the experience of its predecessor
\href{https://espressomd.org}{ESPResSo}, but provides a cleaner, more
modular codebase and enhanced extensibility. It is actively developed by
an international community of researchers across multiple institutions
and disciplines.

\section{Statement of need}\label{statement-of-need}

Molecular dynamics simulations are essential tools for exploring the
behavior of soft matter systems at mesoscopic scales. General-purpose MD
codes such as GROMACS (\citeproc{ref-Abraham2015}{Abraham et al., 2015})
and LAMMPS (\citeproc{ref-Thompson2022}{Thompson et al., 2022}) are
primarily optimized for all-atom simulations and may require significant
customization for coarse-grained models that need non-standard
interactions or specialized multiscale algorithms. ESPResSo\texttt{++}
addresses this gap by providing:

\begin{itemize}
\tightlist
\item
  A modular and extensible design, enabling researchers to easily
  implement new interaction potentials and integrators.
\item
  Efficient parallelization for large-scale simulations of complex
  systems.
\item
  A rich library of coarse-grained interaction models and algorithms
  tailored to soft matter.
\item
  A Python-based scripting interface for ease of use, reproducibility,
  and coupling with external analysis tools.
\item
  Multi-scale simulation techniques such as AdResS and Lees-Edwards.
\end{itemize}

\section{State of the field}\label{state-of-the-field}

Several established MD packages serve the soft matter and coarse-grained
simulation communities. \textbf{LAMMPS}
(\citeproc{ref-Thompson2022}{Thompson et al., 2022}) is a
general-purpose, highly scalable code with broad force field support and
a flexible plugin architecture; however, its input scripting language is
less suited to complex CG workflows, and adaptive resolution is not a
core feature. \textbf{GROMACS} (\citeproc{ref-Abraham2015}{Abraham et
al., 2015}) provides exceptional performance for standard force fields
and supports the Martini CG model ecosystem, but adding truly novel CG
interactions requires modifying the C\texttt{++} source, and its
file-based workflow is less interactive than a Python API.
\textbf{HOOMD-blue} (\citeproc{ref-Anderson2020}{Anderson et al., 2020})
is the most directly comparable package, offering a Python-native API
and GPU-accelerated CG simulations for soft matter; however, it lacks
support for adaptive resolution methods. \textbf{OpenMM}
(\citeproc{ref-Eastman2017}{Eastman et al., 2017}) excels at
GPU-accelerated simulations and allows custom force definitions via
algebraic expressions, but targets primarily biomolecular systems and
lacks MPI-based multi-node parallelization. The sibling project
\textbf{ESPResSo} (\citeproc{ref-Weik2019}{Weik et al., 2019}) shares
common roots but focuses on charged systems, hydrodynamic coupling
(lattice Boltzmann), and electrokinetics rather than multiscale
resolution bridging.

ESPResSo\texttt{++} occupies a distinct niche as a package purpose-built
for coarse-grained and multiscale soft matter simulations. Its key
differentiator is first-class support for adaptive resolution simulation
(AdResS), which allows seamless coupling of atomistic and coarse-grained
representations within a single simulation. ESPResSo\texttt{++} is
currently the only actively maintained MD package that provides AdResS
as a core feature; none of the other packages listed above offer this
capability. It is worth noting that these capabilities were also
contributed to existing general-purpose codes
(\citeproc{ref-Fritsch2012}{Fritsch et al., 2012};
\citeproc{ref-Junghans2010}{Junghans \& Poblete, 2010};
\citeproc{ref-Nagarajan2013}{Nagarajan et al., 2013}), but got removed
again later, hence a dedicated package allows tighter integration of
multiscale algorithms with the domain decomposition, neighbor list
construction, and force computation pipeline --- design choices that
would be difficult to retrofit into architectures optimized for all-atom
throughput.

\section{Software design}\label{software-design}

ESPResSo\texttt{++} uses a layered C\texttt{++}/Python architecture. The
performance-critical computational kernel (force calculations,
integration, domain decomposition, and communication) is implemented in
C\texttt{++}17 and exposed to Python through Boost.Python bindings. Each
Python-exposed C\texttt{++} class provides a static
\texttt{registerPython()} method, and a central registration dispatcher
ensures that the full object hierarchy is available as the
\texttt{espressopp} Python module. This design lets users assemble,
configure, and control simulations entirely from Python while retaining
the performance of compiled C\texttt{++} for the inner loops.

\textbf{Modularity through templates and extensions.}\\
Interactions are implemented using C\texttt{++} class templates
parameterized by a potential type (e.g.,
\texttt{VerletListInteractionTemplate\textless{}\_Potential\textgreater{}}),
so that adding a new pairwise potential requires only implementing the
\texttt{computeEnergy()} and \texttt{computeForce()} methods of a
\texttt{Potential} subclass. The integrator follows a similar pattern:
an \texttt{MDIntegrator} base class exposes an \texttt{addExtension()}
mechanism through which thermostats, barostats, constraints, and
adaptive resolution layers are attached via Boost.Signals2 callbacks.
This signal-based coupling keeps extensions decoupled from the
integration loop and from each other.

\textbf{Parallelization.}\\
ESPResSo\texttt{++} employs spatial domain decomposition with MPI. The
simulation box is partitioned across MPI ranks using a
\texttt{NodeGrid}, and each rank further subdivides its domain into
linked cells via a \texttt{CellGrid}. Ghost particle exchange handles
communication of boundary data. A non-blocking variant
(\texttt{DomainDecompositionNonBlocking}) overlaps communication with
computation, and the heterogeneous spatial domain decomposition
algorithm (HeSpaDDA) (\citeproc{ref-Guzman2017}{Guzman et al., 2017})
provides load balancing for spatially inhomogeneous systems.

\textbf{Performance optimizations.}\\
Since version 2.0 (\citeproc{ref-Guzman2019}{Guzman et al., 2019}),
ESPResSo\texttt{++} has been modernized with SIMD vectorization through
a structure-of-arrays (SOA) particle data layout
(\texttt{ParticleArray}) with 64-byte alignment, yielding an overall
three-times speedup for short-range non-bonded force calculations
(\citeproc{ref-Vance:2023}{Vance et al., 2023}). An improved cell
decomposition scheme (\citeproc{ref-Yao:2004}{Yao et al., 2004}) allows
sub-decomposition into cells with a length of half or a third of the
cutoff radius, reducing the number of unnecessary distance calculations.

\textbf{Testing and documentation.}\\
The code is tested through a combination of Boost.Test (C\texttt{++})
and Python \texttt{unittest} test suites, executed via CMake/CTest and
continuous integration on GitHub Actions. User documentation is built
with Sphinx; developer API documentation with Doxygen.

\section{Research impact statement}\label{research-impact-statement}

ESPResSo\texttt{++} has enabled research across a broad range of soft
matter topics over more than a decade of active development. It has been
used in numerous peer-reviewed studies, including investigations of:

\begin{itemize}
\tightlist
\item
  Polymer rheology and entanglement effects
  (\citeproc{ref-Grommes2021}{Grommes et al., 2021},
  \citeproc{ref-Grommes2022}{2022}, \citeproc{ref-Grommes2024}{2024},
  \citeproc{ref-Grommes:2025}{2025}; \citeproc{ref-Grommes2020}{Grommes
  \& Reith, 2020}; \citeproc{ref-Hsu2020}{Hsu \& Kremer, 2020},
  \citeproc{ref-Hsu2023}{2023}, \citeproc{ref-Hsu2024}{2024};
  \citeproc{ref-Lee2020}{Lee \& Paul, 2020};
  \citeproc{ref-Ohkuma2023}{Ohkuma et al., 2023};
  \citeproc{ref-Singh2020}{Singh et al., 2020};
  \citeproc{ref-Tubiana2021}{Tubiana et al., 2021};
  \citeproc{ref-Zhao2020b}{Zhao, Singh, et al., 2020})
\item
  Polymer concepts in Polysomes organization
  (\citeproc{ref-Guzman2026}{Kobayashi \& Guzman, 2026})
\item
  Lipid membranes, protein and vesicle dynamics
  (\citeproc{ref-Bause2021}{Bause \& Bereau, 2021};
  \citeproc{ref-Pape2023}{Papež et al., 2023};
  \citeproc{ref-Zhao2020}{Zhao, Cortes-Huerto, et al., 2020})
\item
  Adaptive resolution simulations (\citeproc{ref-Bevc2013}{Bevc et al.,
  2013}; \citeproc{ref-Fiorentini2020}{Fiorentini et al., 2020};
  \citeproc{ref-Thaler2020}{Thaler et al., 2020})
\item
  Ionic liquids under shear flow (\citeproc{ref-Gholami2025}{Gholami et
  al., 2025}; \citeproc{ref-Zhang2021}{Zhang et al., 2021})
\item
  Coarse-grained conformational surface hopping
  (\citeproc{ref-Rudzinski2020}{Rudzinski \& Bereau, 2020})
\end{itemize}

The project is developed across multiple institutions --- Max Planck
Institute for Polymer Research, Max Planck Computing and Data Facility,
Los Alamos National Laboratory, Johannes Gutenberg University of Mainz,
and Heidelberg University --- with contributions from both current and
former developers documented in the AUTHORS file. The codebase has a
public development history spanning more than ten years on GitHub, with
continuous integration, code coverage tracking, and versioned releases.

\section{Functionality}\label{functionality}

Key features of ESPResSo\texttt{++} include:

\begin{itemize}
\tightlist
\item
  \textbf{Inter-particle interactions}: Lennard-Jones, Coulomb, soft
  repulsive, bonded interactions, tabulated potentials, and more.
\item
  \textbf{Algorithms}: Molecular dynamics, Langevin dynamics,
  dissipative particle dynamics (DPD), Brownian dynamics, adaptive
  resolution simulations (AdResS), Monte Carlo sampling.
\item
  \textbf{Electrostatics}: Particle--particle particle--mesh (P3M),
  Ewald summation, and other long-range methods.
\item
  \textbf{Parallelization}: Domain decomposition using MPI, optimized
  for massively parallel architectures.
\item
  \textbf{Python interface}: Full simulation control and analysis
  scripting in Python.
\item
  \textbf{Extensibility}: Modular design allows easy addition of new
  force fields, integrators, or analysis tools.
\end{itemize}

\section{\texorpdfstring{New Features since ESPResSo\texttt{++}
v2.0}{New Features since ESPResSo v2.0}}\label{new-features-since-espresso-v2.0}

Since the last major release of ESPResSo\texttt{++} v2.0 in 2018
(\citeproc{ref-Guzman2019}{Guzman et al., 2019}) a number of new
functionalities and features have been added, including:

\begin{itemize}
\tightlist
\item
  \textbf{SIMD vectorization and related optimizations}: enhance compute
  performance on modern CPUs (\citeproc{ref-Vance:2023}{Vance et al.,
  2023})
\item
  \textbf{Cell decomposition}: allow sub-decomposition into cells with a
  length of half or a third of the cutoff for direct force calculations
  (\citeproc{ref-Yao:2004}{Yao et al., 2004})
\item
  \textbf{HeSpaDDA}: heterogeneous spatial domain decomposition
  algorithm (HeSpaDDA) for larger scale simulations
  (\citeproc{ref-Guzman2017}{Guzman et al., 2017})
\item
  \textbf{new potentials and simulation methods}: AngularCosineSquared,
  TabulatedSubEnsAngular, surface hopping MD, Lees-Edwards boundary
  conditions
\item
  \textbf{Checkpoint the state of the random number generator (RNG)}:
  allow restarting from checkpointed state of RNG
\item
  \textbf{I/O}: support for parallel writing and reading of H5MD
  checkpoints
\item
  \textbf{Python 3 compatibility}
\end{itemize}

\section{Example usage}\label{example-usage}

A minimal Python script to run a simple (repulsive only) Lennard-Jones
type particle simulation in ESPResSo\texttt{++} looks like:

\begin{Shaded}
\begin{Highlighting}[]
\ImportTok{import}\NormalTok{ espressopp}
\ImportTok{from}\NormalTok{ mpi4py }\ImportTok{import}\NormalTok{ MPI}

\CommentTok{\# simulation system parameters}
\NormalTok{num\_particles }\OperatorTok{=} \DecValTok{10000}      \CommentTok{\# total number of particles in the system}
\NormalTok{box           }\OperatorTok{=}\NormalTok{ (}\DecValTok{20}\NormalTok{,}\DecValTok{20}\NormalTok{,}\DecValTok{20}\NormalTok{) }\CommentTok{\# size of the simulationbox (all length are in sigma)}
\NormalTok{rc            }\OperatorTok{=} \FloatTok{1.12246}    \CommentTok{\# cut off for the short range non bonded potential}
\NormalTok{skin          }\OperatorTok{=} \FloatTok{0.3}        \CommentTok{\# skin used for verlet neighbor list}
\NormalTok{dt            }\OperatorTok{=} \FloatTok{0.005}      \CommentTok{\# time step for 1 md step}
\NormalTok{epsilon       }\OperatorTok{=} \FloatTok{1.0}        \CommentTok{\# energy unit}
\NormalTok{sigma         }\OperatorTok{=} \FloatTok{1.0}        \CommentTok{\# length unit}
\NormalTok{temperature   }\OperatorTok{=} \FloatTok{1.0}        \CommentTok{\# temperature of the simulation}
\NormalTok{LJcaprad      }\OperatorTok{=} \FloatTok{0.8}        \CommentTok{\# inital capping radius for LJ interaction}
                           \CommentTok{\# for random configurations}

\CommentTok{\# system setup}
\NormalTok{system         }\OperatorTok{=}\NormalTok{ espressopp.System()}
\NormalTok{system.rng     }\OperatorTok{=}\NormalTok{ espressopp.esutil.RNG()}
\NormalTok{system.bc      }\OperatorTok{=}\NormalTok{ espressopp.bc.OrthorhombicBC(system.rng, box)}
\NormalTok{system.skin    }\OperatorTok{=}\NormalTok{ skin}

\CommentTok{\# define underlying storage system for parallelisation}
\NormalTok{nodeGrid       }\OperatorTok{=}\NormalTok{ espressopp.tools.decomp.nodeGrid(MPI.COMM\_WORLD.size,box,rc,skin)}
\NormalTok{cellGrid       }\OperatorTok{=}\NormalTok{ espressopp.tools.decomp.cellGrid(box, nodeGrid, rc, skin)}
\NormalTok{system.storage }\OperatorTok{=}\NormalTok{ espressopp.storage.DomainDecomposition(system, nodeGrid, cellGrid)}

\CommentTok{\# interaction setup, here short range non{-}bonded Lennard Jones potential}
\NormalTok{interaction    }\OperatorTok{=}\NormalTok{ espressopp.interaction.VerletListLennardJonesCapped(}
\NormalTok{                   espressopp.VerletList(system, cutoff}\OperatorTok{=}\NormalTok{rc))}
\NormalTok{interaction.setPotential(type1}\OperatorTok{=}\DecValTok{0}\NormalTok{, type2}\OperatorTok{=}\DecValTok{0}\NormalTok{,}
\NormalTok{                         potential}\OperatorTok{=}\NormalTok{espressopp.interaction.LennardJonesCapped(}
\NormalTok{                           epsilon, sigma, rc, shift}\OperatorTok{=}\StringTok{\textquotesingle{}auto\textquotesingle{}}\NormalTok{, caprad}\OperatorTok{=}\NormalTok{LJcaprad))}
\NormalTok{system.addInteraction(interaction)}

\CommentTok{\# integrator setup}
\NormalTok{integrator     }\OperatorTok{=}\NormalTok{ espressopp.integrator.VelocityVerlet(system)}

\CommentTok{\# thermostat setup}
\NormalTok{thermostat             }\OperatorTok{=}\NormalTok{ espressopp.integrator.LangevinThermostat(system)}
\NormalTok{thermostat.gamma       }\OperatorTok{=} \FloatTok{1.0}
\NormalTok{thermostat.temperature }\OperatorTok{=}\NormalTok{ temperature}
\NormalTok{integrator.addExtension(thermostat)}

\CommentTok{\# create random particle setup in the simulation box}
\NormalTok{props }\OperatorTok{=}\NormalTok{ [}\StringTok{\textquotesingle{}id\textquotesingle{}}\NormalTok{, }\StringTok{\textquotesingle{}type\textquotesingle{}}\NormalTok{, }\StringTok{\textquotesingle{}mass\textquotesingle{}}\NormalTok{, }\StringTok{\textquotesingle{}pos\textquotesingle{}}\NormalTok{, }\StringTok{\textquotesingle{}v\textquotesingle{}}\NormalTok{]}
\NormalTok{new\_particles }\OperatorTok{=}\NormalTok{ []}
\NormalTok{pid }\OperatorTok{=} \DecValTok{1}
\ControlFlowTok{while}\NormalTok{ pid }\OperatorTok{\textless{}=}\NormalTok{ num\_particles:}
    \BuiltInTok{type} \OperatorTok{=} \DecValTok{0}
\NormalTok{    mass }\OperatorTok{=} \FloatTok{1.0}
\NormalTok{    pos  }\OperatorTok{=}\NormalTok{ system.bc.getRandomPos()}
\NormalTok{    vel  }\OperatorTok{=}\NormalTok{ espressopp.Real3D(}\FloatTok{0.0}\NormalTok{, }\FloatTok{0.0}\NormalTok{, }\FloatTok{0.0}\NormalTok{)}
\NormalTok{    part }\OperatorTok{=}\NormalTok{ [pid, }\BuiltInTok{type}\NormalTok{, mass, pos, vel]}
\NormalTok{    new\_particles.append(part)}
    \ControlFlowTok{if}\NormalTok{ pid }\OperatorTok{\%} \DecValTok{1000} \OperatorTok{==} \DecValTok{0}\NormalTok{:}
\NormalTok{        system.storage.addParticles(new\_particles, }\OperatorTok{*}\NormalTok{props)}
\NormalTok{        system.storage.decompose()}
\NormalTok{        new\_particles }\OperatorTok{=}\NormalTok{ []}
\NormalTok{    pid }\OperatorTok{+=} \DecValTok{1}
\NormalTok{system.storage.addParticles(new\_particles, }\OperatorTok{*}\NormalTok{props)}

\NormalTok{integrator.dt }\OperatorTok{=} \FloatTok{0.0001} \CommentTok{\# very small timestep for initial warmup}
\ControlFlowTok{for}\NormalTok{ n }\KeywordTok{in} \BuiltInTok{range}\NormalTok{(}\DecValTok{20}\NormalTok{):}
  \CommentTok{\# warmup finished, switch to uncapped Lennard Jones potential and increase timestep dt}
  \ControlFlowTok{if}\NormalTok{ n }\OperatorTok{==} \DecValTok{10}\NormalTok{: }
\NormalTok{    interaction }\OperatorTok{=}\NormalTok{ espressopp.interaction.VerletListLennardJones(}
\NormalTok{                    espressopp.VerletList(system, cutoff}\OperatorTok{=}\NormalTok{rc))}
\NormalTok{    interaction.setPotential(type1}\OperatorTok{=}\DecValTok{0}\NormalTok{, type2}\OperatorTok{=}\DecValTok{0}\NormalTok{,}
\NormalTok{                             potential}\OperatorTok{=}\NormalTok{espressopp.interaction.LennardJones(}
\NormalTok{                               epsilon, sigma, rc, shift}\OperatorTok{=}\StringTok{\textquotesingle{}auto\textquotesingle{}}\NormalTok{))}
\NormalTok{    system.removeInteraction(}\DecValTok{0}\NormalTok{)}
\NormalTok{    system.addInteraction(interaction)}
\NormalTok{    integrator.dt }\OperatorTok{=}\NormalTok{ dt}
\NormalTok{  integrator.run(}\DecValTok{10000}\NormalTok{)}
\NormalTok{  Etot }\OperatorTok{=}\NormalTok{ system.getInteraction(}\DecValTok{0}\NormalTok{).computeEnergy()}
  \BuiltInTok{print}\NormalTok{(}\StringTok{"md time = }\SpecialCharTok{\{:4.1f\}}\StringTok{, total energy: }\SpecialCharTok{\{:10.2f\}}\StringTok{"}\NormalTok{.}\BuiltInTok{format}\NormalTok{(integrator.dt}\OperatorTok{*}\NormalTok{n}\OperatorTok{*}\DecValTok{10000}\NormalTok{, Etot))}

\CommentTok{\# write PDB file of (quasi) equilibrated LJ system.}
\CommentTok{\# At this temperature it is more or less crystallized.}
\NormalTok{espressopp.tools.pdbwrite(}\StringTok{"simplelj.pdb"}\NormalTok{, system, molsize}\OperatorTok{=}\NormalTok{num\_particles)}
\end{Highlighting}
\end{Shaded}

\section{AI usage disclosure}\label{ai-usage-disclosure}

No generative AI tools were used in the creation of the
ESPResSo\texttt{++} software or its documentation. During the
preparation of this manuscript, AI-assisted tools were used for
copyediting and formatting. All content was reviewed and verified by the
authors.

\section{Acknowledgements}\label{acknowledgements}

We thank the ESPResSo\texttt{++} developer community and all
contributors listed in the AUTHORS file. ESPResSo\texttt{++} has been
supported by the Transregio TRR146 of the German Research Foundation.
The ESPResSo\texttt{++} project is supported by the U.S. Department of
Energy through Los Alamos National Laboratory (LANL). Los Alamos
National Laboratory is operated by Triad National Security, LLC, for the
National Nuclear Security Administration of the U.S. Department of
Energy (contract no. 89233218CNA000001). This paper has been assigned a
Los Alamos Unlimited Release number of LA-UR-26-21700.

\protect\phantomsection\label{refs}

\end{document}